\newcommand{\beq}{\begin{equation}}
\newcommand{\eeq}{\end{equation}}
\newcommand{\eq}[1]{\begin{align}#1\end{align}}

\documentclass[prl,letterpaper,aps,floatfix,twocolumn]{revtex4}
\usepackage[dvipdfmx]{graphicx} 
\usepackage{bm}
\usepackage{booktabs}
\usepackage{amsmath,amsthm,amssymb}
\usepackage{color}

\usepackage{ascmac}

\begin{document}

\title{Photovoltaic Chiral Magnetic Effect}
\author{Katsuhisa Taguchi$^{1,3}$, Tatsushi Imaeda$^{1,3}$, Masatoshi Sato$^{2}$, and Yukio Tanaka$^{1,3}$}
\affiliation{$^1$Department of Applied Physics, Nagoya University, Nagoya, 464-8603, Japan
\\
$^2$Yukawa Institute for Theoretical Physics, Kyoto University, Kyoto, 606-8502, Japan 
\\
$^3$CREST, Japan Science and Technology Corporation (JST), Nagoya 464-8603, Japan 
}
\date{\today } 
%

\begin{abstract}
We theoretically predict a generation of a current in Weyl semimetals 
by applying circularly polarized light.  
The electric field of the light can drive an
effective magnetic field of order of ten Tesla.
For lower frequency light, a non-equilibrium spin distribution is
formed near the Fermi surface. 
Due to the spin-momentum locking, 
a giant electric  current proportional to the effective magnetic field is
 induced. 
On the other hand, higher frequency light realizes a quasi-static
 Floquet state with no induced electric current. 
We  discuss relevant materials and estimate order of magnitude of the
induced current.
 

%

\end{abstract}

\maketitle


{\bf Introduction}---
Recently, Dirac and Weyl semimetals, 
which host bulk gapless excitations obeying quasi-relativistic fermion
equations, have attracted much attention in condensed matter physics  
\cite{rf:Murakami07, rf:burkov11,rf:Yang14,rf:biinse1,rf:biinse2, rf:nabi1, rf:nabi2, rf:cd2as3theo, rf:cd2as3exp, rf:taas1, rf:taas2,rf:taas3,rf:taas4, rf:tominaga14}. 
Dirac semimetals 
have been theoretically predicted\cite{rf:Murakami07, rf:burkov11,rf:Yang14} 
and experimentally demonstrated 
in (Bi$_{1-x}$In$_{x}$)$_{2}$Se$_{3}$\cite{rf:biinse1, rf:biinse2},  
Na$_{3}$Bi\cite{rf:nabi1, rf:nabi2} and
Cd$_{3}$As$_{2}$\cite{rf:cd2as3theo, rf:cd2as3exp}.
There are also several experiments   
supporting the realization of Weyl semimetals 
in TaAs\cite{rf:taas1, rf:taas2, rf:taas3, rf:taas4}. 
Moreover, Dirac and Weyl semimetals have been theoretically predicted
in a superlattice heterostructure of  topological insulator (TI)/normal insulator
(NI)\cite{rf:burkov11}, and a Dirac semimetal has been realized in the
GeTe/Sb$_2$Te$_3$ superlattice\cite{rf:tominaga14}. 

%

Low energy bulk excitations in Dirac and Weyl semimetals come in pairs
of left and right-handed Weyl fermions,
because of the Nielsen-Ninomiya's no go theorem\cite{rf:Nielsen81}. 
In the low energy limit, each charge flow of left and right-handed Weyl
fermions preserves classically, but their difference,
the axial current, is not conserved in the quantum theory, due to the
chiral anomaly.
In an analogy of relativistic high energy physics
\cite{rf:Vilenkin80,rf:Metlitski05, rf:Kharzeev08, rf:Fukushima08, rf:Kharzeev13,rf:Kharzeev14},
the anomaly related effects have been discussed in condensed matter
physics\cite{rf:Jackiw99,rf:Vazifeh13,rf:Yamamoto12,rf:Zyuzin12,rf:Zyuzin12b,rf:Chen13,rf:Burkov15,
rf:Sumiyoshi15,rf:Nomura15,rf:Taguchi15}.
The anomaly induced currents are dissipationless and thus they  
have potential applications to unique electronics. 


%


Among the anomaly related effects, one of the most interesting phenomena is
the chiral magnetic effect.
In the presence of a time-dependent $\theta$ term in the Dirac-Weyl theory, 
a current proportional to an applied static magnetic field has been predicted
theoretically\cite{rf:Vilenkin80,rf:Metlitski05,rf:Kharzeev08,rf:Fukushima08,rf:Kharzeev13,rf:Zyuzin12,rf:Zyuzin12b,rf:Vazifeh13,rf:Yamamoto12}. 
The flow due to the static magnetic field, however, might
be problematic in condensed matter physics.
First, in Weyl semimetals, the time-dependent $\theta$ term is
obtained in the ground state, in the presence of the energy difference of
left and right-handed Weyl points \cite{rf:Zyuzin12b}. 
%
However, the system stays the ground state under a static magnetic field, so no
actual current should flow eventually\cite{rf:Vazifeh13}. 
Moreover, the detection can be difficult because there is no driving force
to get out the current in such an equilibrium state. 
Hence, instead of a static magnetic field, one should consider a
non-equilibrium magnetic field to obtain a net current of the chiral
magnetic effect.


Recent studies using femtosecond laser pulses have established
a method to generate  
non-equilibrium magnetic fields by circularly polarized light in ferrimagnets\cite{rf:Stanciu07,rf:Vahaplar09,rf:Kirilyuk10}.
The light-induced effective magnetic field 
$\bm{B}^\textrm{eff}$ is given by
%
\begin{eqnarray}
\bm{B}^\textrm{eff} \propto i\bm{\mathcal{E}} \times \bm{\mathcal{E}}^*.
\label{eq:1-1} 
\end{eqnarray}
with the circularly polarized complex electric field $\bm{\mathcal{E}}$\cite{rf:Pitaevskii61, rf:Pershan66}.
The generation of the effective magnetic field is due to the conversion of 
spin-angular momentum from light to electrons via the spin-orbit
coupling\cite{rf:Pershan66,rf:Taguchi11,rf:Misawa11}.
The direction of $\bm{B}^\textrm{eff}$ depends on the chirality of 
the circularly polarized light. 
Its magnitude is proportional 
to the intensity of laser and 
can be 20 T for a sufficient strong laser pulse\cite{rf:Stanciu07,rf:Vahaplar09,rf:Kirilyuk10}.

\begin{figure}[b]\centering 
\includegraphics[scale=.37]{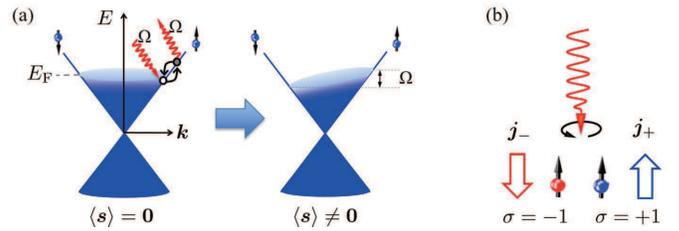} 
\caption{(Color online) 
Schematic illustration of photovoltaic chiral magnetic effect: 
(a) For a lower frequency light regime, 
electrons near the Fermi surface are excited by the incident light
 through the Raman process illustrated.
As a result, a finite spin distribution is generated near the Fermi
 surface, and the spin of
 Weyl fermions is aligned in the direction of the effective magnetic
 field ${\bm B}^{\rm eff}$, on average.
(b) 
For the reason above, 
the circularly polarized light 
aligns the spin of Weyl fermions. 
Because of (pseudo)spin-momentum locking, Weyl fermions with helicity
 $\sigma=1$ $(\sigma=-1)$ move in the same (opposite) direction as the spin, 
which results in nonzero current $\bm{j}_\sigma$. 
}
\label{fig:fig1} 
\end{figure}

In this Letter, we theoretically predict a current $\bm
 {j}$ induced by the effective magnetic field (Fig.\ref{fig:fig1}).
The photovoltaic current is due to a non-equilibrium spin
distribution near the Fermi surface.
For lower frequency light, the conversion of spin-angular momentum
between light and electrons occurs only near the Fermi surface.
Thus, like the relativistic theory, the low energy description using
Weyl fermions gives a good approximation to evaluate the chiral
magnetic effect.
On the basis of the Keldysh Green's function, we show that a net current
is obtained by applying circularly polarized light.
The current is proportional to the effective magnetic field in an analogous form
of the chiral magnetic effect.
On the other hand, when light has a frequency higher than the energy
scale of the band width, a  quasi-static Floquet state is realized, 
where the chiral magnetic effect is cancelled due to the occupied band electrons.
In the latter case, Weyl points are shifted in the momentum
space, resulting in the change of the anomalous Hall effect, instead.
%


{\bf Model}---
We consider 
the following Hamiltonian to describe
Weyl/Dirac semimetals in the presence of circularly polarized light
\eq{\label{eq:2-1} 
H & = H_{\textrm{Weyl}} + H_{\textrm{em}} + V_{\textrm{imp}}.
}
The first term is the Hamiltonian of Weyl/Dirac semimetals.
In low energy, it takes the form
\eq{
	\label{eq:2-2} 
H_{\textrm{Weyl}} & =  \sum_{\bm{k}}  \psi^\dagger_{\bm{k}} 
	\mathcal{H}_{\textrm{Weyl}} \psi_{\bm{k}},\\
\mathcal{H}_{\textrm{Weyl}} & =  
	\hbar v_{\textrm{F}} \sigma^z (\bm{k}  - \sigma^z \bm{b})
\cdot\bm{s} 
	- \mu\sigma^0s^0-\mu_5\sigma^z s^0,
}
where  $\psi_{\bm{k}} = {}^t\!(\psi_{\uparrow, +} \  \psi_{\downarrow, +} \ \psi_{\uparrow, -} \  \psi_{\downarrow, -} )$ 
is the annihilation operator of electron with (pseudo)spin
($\uparrow,\downarrow$) and helicity ($+, -$). 
$s^\mu$ and $\sigma^\mu$ are the Pauli matrices of (pseudo)spin and
helicity, $v_{\textrm{F}}$ is the Fermi velocity, and $\mu$
is the chemical
potential.
The parameters $2\bm{b}$ and $2\mu_5$ denote the difference of the position
of left and right-handed Weyl points in the momentum and energy spaces,
respectively.
For Dirac semimetals, $\bm{b}=\bm{0}$ and $\mu_5=0$.
The second term in Eq. (\ref{eq:2-2}) is the gauge coupling between
Weyl/Dirac semimetals and light 
\eq{
\label{eq:2-3} 
H_{\textrm{em}} & = - \sum_{\bm{k}}  \bm{j} \cdot \bm{A}^\textrm{em}, 
}
where $\bm{j}$ denotes the charge current, 
and $\bm{A}^\textrm{em}$ is the vector potential of light.
For circlarily polarized light, the electric field $\bm{E}^\textrm{em} =
-\partial_t \bm{A}^\textrm{em}$ is 
given by
\eq{ 
	\label{eq:2-4} 
\bm{E}^\textrm{em} 
	& =  \textrm{Re}\left[ \bm{\mathcal{E}} e^{i\Omega t} \right],
}
where $\bm{\mathcal{E}}$ is a complex vector and 
$\Omega $ is the angular frequency of light.
%
The third term in Eq. (\ref{eq:2-1}) expresses the  
impurity scattering in Weyl/Dirac semimetals\cite{rf:book1,rf:Hosur12},
\eq{\label{eq:2-5} 
V_{\textrm{imp}}
	& =  \sum_{\bm{k},\bm{q}} \psi^\dagger_{\bm{k}+\bm{q}} 
	\sigma^0  s^0 u_{\textrm{imp}} (\bm{q}) \psi_{\bm{k}}. 
}
The impurity scattering potential $u_{\textrm{imp}} $ is 
assumed to be short-ranged
and triggers a finite relaxation time, 
which is given within the Born approximation as 
$\tau_{{\rm e},\sigma} = \hbar/(\pi  \nu_{{\rm e},\sigma} n_\textrm{c} u_{\textrm{imp}}^2)$
with a concentration of nonmagnetic impurities $n_\textrm{c}$.

{\bf Current induced by circularly polarized light}---
We calculate the current induced by light, 
using the Keldysh Green's function technique\cite{rf:book1}.
Below, we assume that $\hbar\Omega$ is much lower than the
band width, so the low energy effective Hamiltonian (\ref{eq:2-2}) gives a
good approximation. 
For Eq.(\ref{eq:2-2}),
the current is defined by 
\eq{ \label{eq:3-1a} 
\langle \bm{j}\rangle 
	 \equiv ev_\textrm{F} \langle \psi^\dagger(\bm{x},t) \sigma^z
	 \bm{s} \psi(\bm{x},t) \rangle, 
}
which is decomposed as
\eq{ \label{eq:3-1b} 
\langle \bm{j} \rangle 
	 \equiv \langle \bm{j}_{+} \rangle +\langle \bm{j}_{-} \rangle,
}
with $\langle \bm{j}_{\sigma=\pm} \rangle  
	\equiv  \sigma ev_\textrm{F} \langle \psi^\dagger_{\sigma}
	\bm{s} \psi_{\sigma} \rangle$.
Here $\psi^\dagger_{\sigma}=(\psi^\dagger_{\sigma,\uparrow},
\psi^\dagger_{\sigma,\downarrow})$ is the creation operator of Weyl
fermions with helicity $\sigma=\pm$.
There is no mixing term between $\psi^{\dagger}_+$
and $\psi_-$ in $H$, and thus
$\langle {\bm j}_+ \rangle$ and $\langle {\bm j}_- \rangle$ can be
calculated separately.
%
For a while, we consider the  ${\bm b}=0$ case.

\begin{figure}[h]\centering
\includegraphics[scale=.22]{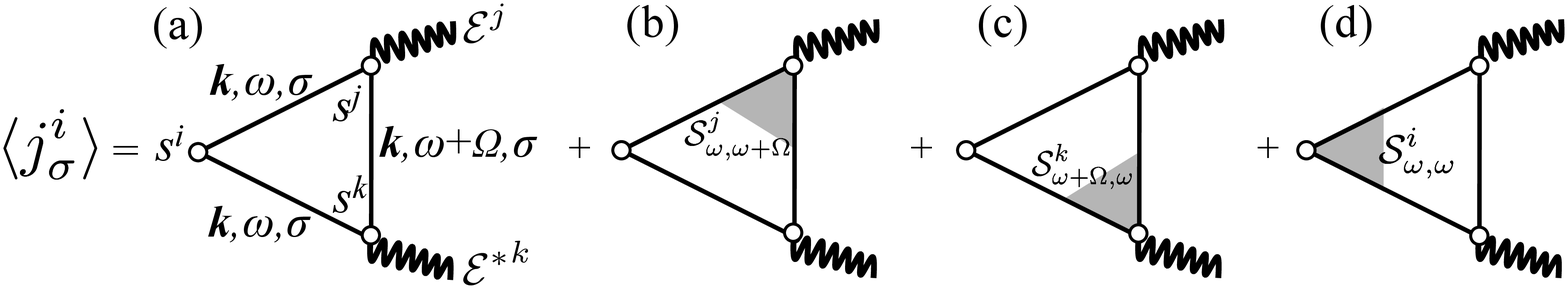}  
\caption{
Diagrammatic representation of a charge current $\langle
 j^i_\sigma \rangle $ via photovoltaic
chiral magnetic effect 
(a) without and (b-d) with a vertex correction of
 nonmagnetic impurity scattering. 
 A wavy line denotes an electric
 field of the polarized light.
 }
\label{fig:diagram} 
\end{figure}

In terms of the Keldysh Green's function, the chiral current $\langle
\bm{j}_{\sigma} \rangle$ is represented as 
$\langle \bm{j}_{\sigma} \rangle 
	 = -\sigma i \hbar ev_\textrm{F} 
\textrm{tr} [ \bm{s} G^<_{\sigma} (\bm{x},t:\bm{x},t) ]$  
with the $2\times2$ matrix lesser Green function $G_\sigma^<
(\bm{x},t:\bm{x},t) = - i\hbar \langle \psi^\dagger_\sigma (\bm{x},t)
\psi_\sigma (\bm{x},t) \rangle $.
The contribution from $\bm{B}^\textrm{eff} \propto i\bm{\mathcal{E}}
\times \bm{\mathcal{E}^*}$ is given by the diagrams in Fig.\ref{fig:diagram}.
It is written as
\begin{eqnarray}
\langle j^i_{\sigma}\rangle= -i\hbar ev_{\rm F}
[{\cal I}_{\sigma}^{ijk}(\Omega)
+{\cal I}_{\sigma}^{ijk}(-\Omega)
]{\cal E}^j{{\cal E}^*}^k
\end{eqnarray}
with
\begin{eqnarray}
{\cal I}_{\sigma}^{ijk}(\Omega)
=\frac{e^2v_{\rm F}^2}{4\Omega^2}
\sum_{I=a,b,c,d}
{\cal C}_\sigma^{(I), ijk}.
\end{eqnarray}
Each diagram in Fig.\ref{fig:diagram} gives the following ${\cal
C}_\sigma^{(I=a,b,c,d), ijk}$ 
\begin{eqnarray}
&&{\cal C}_\sigma^{(a), ijk}=\sum_{{\bm k},\omega}{\rm tr}
\left[s^i g_{{\bm k},\omega,\sigma}s^jg_{{\bm k}, \omega+\Omega,\sigma}
s^k g_{{\bm k}, \omega,\sigma}
\right]^< ,
\nonumber\\
&&{\cal C}_\sigma^{(b), ijk}=\sum_{{\bm k},\omega}{\rm tr}
\left[s^i g_{{\bm k},\omega,\sigma}{\cal S}^j_{\omega,\omega+\Omega}g_{{\bm k}, \omega+\Omega,\sigma}
s^k g_{{\bm k}, \omega,\sigma}
\right]^<, 
\nonumber\\
&&{\cal C}_\sigma^{(c), ijk}=\sum_{{\bm k},\omega}{\rm tr}
\left[s^i g_{{\bm k},\omega,\sigma}s^jg_{{\bm k}, \omega+\Omega,\sigma}
{\cal S}^k_{\omega+\Omega, \omega} g_{{\bm k}, \omega,\sigma}
\right]^<, 
\nonumber\\
&&{\cal C}_\sigma^{(d), ijk}=\sum_{{\bm k},\omega}{\rm tr}
\left[{\cal S}^i_{\omega,\omega} 
g_{{\bm k},\omega,\sigma}s^jg_{{\bm k}, \omega+\Omega,\sigma}
s^k g_{{\bm k}, \omega,\sigma}
\right]^< ,
\end{eqnarray}
where $g_{{\bm k},\omega,\sigma}^<$ is given by
\begin{eqnarray}
g_{{\bm k},\omega,\sigma}^<=f_{\omega}\left[
g_{{\bm k},\omega,\sigma}^{\rm a} 
-g_{{\bm k},\omega,\sigma}^{\rm r} 
\right]
\label{eq:lg}
\end{eqnarray}
with the Fermi distribution function $f_\omega$ and the retarded and
advanced Green's functions 
\begin{eqnarray}
&&g^{\rm r}_{{\bm k},\omega,\sigma}=
\left[\hbar\omega-\sigma\hbar v_{\rm F}{\bm k}\cdot {\bm
 s}+\mu+\sigma\mu_5+\frac{i\hbar}{2\tau_{{\rm e}, \sigma}}
\right]^{-1},
\nonumber\\
&&g^{\rm a}_{{\bm k},\omega,\sigma}= \left[
g^{\rm r}_{{\bm k},\omega,\sigma}\right]^\dagger.
\end{eqnarray}
${\cal S}^i_{\omega,\omega'}$ is the vertex correction due to the
nonmagnetic impurity scattering $V_{\rm imp}$\cite{rf:suppl-A}.  

%
%
%

Using Eq.(\ref{eq:lg}), one can rewrite $C_{\sigma}^{(I=a,b,c,d),ijk}$ in terms
of the retarded and advanced Green's functions.
For $|\mu+\sigma\mu_5|\gg\hbar/\tau_{\rm e}$, we find that
\begin{eqnarray}
C_{\sigma}^{(I=a,b,c,d),ijk}\propto \sum_{\omega}(f_{\omega+\Omega}-f_{\omega}). 
\end{eqnarray}
This means that only fermions near the Fermi surface contribute the
light-induced current, which justifies  our Weyl fermion approximation.
We also find that $C_{\sigma}^{(I),ijk}$ contains both of the
retarded and advanced Green's functions, and it is expressed by their product.
This is a signal of a non-equilibrium process\cite{rf:book1}.

After some calculation \cite{rf:suppl-A}, we obtain  
\begin{eqnarray}
\langle {\bm j}_{\sigma}\rangle =
\sigma \frac{2\nu_{{\rm e},\sigma}e^3v_{\rm F}^3\tau_{{\rm e},\sigma}^4}{3\hbar}
\Omega i
({\bm{\mathcal E}}\times {\bm{\mathcal E}^*}),
\label{eq:3-3}
\end{eqnarray}
where $\nu_{\rm e,\sigma}=\frac{(\mu+\sigma \mu_5)^2}{2\pi^2 \hbar^3 v_\textrm{F}^3 }$ is the density of state of the Weyl cone with
helicity $\sigma$.
From Eq. (\ref{eq:3-3}), the total current $\langle
\bm{j}\rangle $ is
\begin{eqnarray}
\langle {\bm j}\rangle=
\frac{2(\nu_{{\rm e}, +} \tau^4_{{\rm e},+} -\nu_{{\rm e}, -}\tau^4_{{\rm e},-} )e^3v_{\rm F}^3 }{3\hbar}
\Omega i
({\bm{\mathcal E}}\times {\bm{\mathcal E}^*}),
\label{eq:current}
\end{eqnarray}
which is non-zero 
when $\nu_{{\rm e}, +} \tau^4_{{\rm e},+} \neq \nu_{{\rm e}, -}\tau^4_{{\rm
e},-}$, namely when $\mu_5\neq 0$.

The obtained current originates from a non-equilibrium distribution of spin:
When one exposes the system to circularly polarized light,  
the conversion of spin-angular momentum
between light and electrons occurs due to the spin-orbit interaction.
As a result, there arises a non-equilibrium distribution of spin near
the Fermi surface [Fig.\ref{fig:fig1}(a)].
For Weyl fermions, because of the spin-momentum locking, the
non-equilibrium spin distribution gives rise to the current flow
[Fig.\ref{fig:fig1}(b)].
Indeed, for Eq.(\ref{eq:2-2}), 
the current operator is essentially the same as the
spin operator, and thus, from the same calculation, one can show that the circularly
polarized light induces a non-zero spin polarization of electrons 
\begin{eqnarray}
\langle\psi^{\dagger}_{\sigma} {\bm s} \psi_{\sigma}\rangle
= 
\frac{2\nu_{{\rm e},\sigma}e^2v_{\rm F}^2 \tau^4_{{\rm e},\sigma} }{3\hbar}
\Omega i
({\bm{\mathcal E}}\times {\bm{\mathcal E}^*}),
\end{eqnarray}
near the Fermi surface.

Since the circularly polarized light induces the spin-polarization of
electrons, it effectively acts as a Zeeman magnetic field near the Fermi surface
\eq{
	\label{eq:3-4} 
\bm{B}^{\textrm{eff}}_\sigma 
	& \equiv \chi_\sigma \Omega i \bm{\mathcal{E}} \times \bm{\mathcal{E}}^*
	=\sigma_{\textrm{L}} \chi_\sigma \Omega  |\bm{\mathcal{E}}|^2 \hat{\bm{q}},
}
with $\chi_\sigma \equiv \frac{2}{3} \frac{ e^2 v_\textrm{F}^2 \tau^4_{{\rm e},\sigma} }{ g
\mu_\textrm{B} \hbar }$.
Here $\hat{\bm q}$ is the unit vector of the direction of 
light propagation,
$\sigma_{\textrm{L}}=\pm 1$ specifies the chirality (clockwise or
counter-clockwise polarization) of light, $g$ is
the Land\'{e} factor, and $\mu_{\rm B}$ is 
the Bohr magneton.

It is noted that the light-induced current has a similarity to
the chiral magnetic effect. 
In both cases, the current flows in the direction of an applied magnetic
or effective magnetic field, and its magnitude is proportional to the
difference of the chemical potential between left and right-handed fermions.
Indeed, like our case, the spin-polarization and the spin-momentum
locking are essential to obtain the current in the chiral magnetic
effect\cite{rf:Kharzeev08}.
Under a static magnetic field, electrons form the Landau levels. 
For Weyl fermions, the zeroth Landau level is fully
spin-polarized in the direction of the applied magnetic field, and thus
the ground state of the system is also spin-polarized.
As a result, the current flows 
due to the spin-momentum locking\cite{rf:Kharzeev08}. 
We dub our light-induced current effect as photovoltaic chiral magnetic effect.

Here we would like to mention that there is an important difference
between our photovoltaic chiral magnetic effect and the original one.
In the original case, the chiral magnetic effect is caused by a static
magnetic field, and thus the resultant current is
equilibrium (and dissipationless).
In condensed matter physics, however, an analogous current of Weyl
fermions, even if exists, is completely cancelled by other current in
the conduction band \cite{rf:Vazifeh13}.
On the other hand, the photovoltaic chiral magnetic effect is due to the
time-dependent electric field, so the current is non-equilibrium and dissipative.
The current comes only from Weyl fermions near the Fermi surface, so no
cancellation occurs.

The effective magnetic field also generates the axial current, which is
the difference between charge currents with
different helicity: $\langle \bm{j}_\textrm{axial} \rangle \equiv
\langle \bm{j}_{+} \rangle -\langle \bm{j}_{-} \rangle
=ev_\textrm{F} 
[\langle\psi^{\dagger}_{+} {\bm s} \psi_{+}\rangle
+ \langle\psi^{\dagger}_{-} {\bm s} \psi_{-}\rangle]$.
As mentioned above, for lower $\Omega$, the system is well described by
Weyl fermions, and thus the axial current can be well-defined as well.
The axial current is nonzero even for Dirac semimetals with ${\bm
b}=\mu_5=0$.  
%
The axial current can be detected as total spin polarization, by using
pump-probe techniques\cite{rf:Kirilyuk10}.

We can easily generalize the above result for $\langle {\bm j}\rangle$ in the case with ${\bm b}\neq 0$.
Since ${\bm b}$ behaves like a static Zeeman field in ${\cal H}_{\rm Weyl}$, 
it can shift $\langle \psi^{\dagger}_{\sigma}{\bm s}\psi_\sigma\rangle$ by
the Pauli paramagnetism.
However, ${\bm b}$ cannot drive a net current
since it is static.
Moreover, the circularly polarized light affects only electrons near the
Fermi surface, which structure does not depend on ${\bm b}$.  
Therefore, we have the same current $\langle {\bm j}\rangle$ in Eq.(\ref{eq:current})
even when ${\bm b}\neq 0$.

We estimate the magnitude of ${\bm B}^\textrm{eff}_\sigma$ and $\langle \bm{j} \rangle $ by using 
material parameters for TaAs\cite{rf:C-Zhang15}, 
$v_\textrm{F}=3\times 10^5$ m/s,
$\tau_e=4.5\times 10^{-11}$ s, and 
$\mu=11.5$ meV.
If the difference of the chemical potential is $\mu_5=1$ meV,
$|{\bm B}^{\textrm{eff}}_\sigma|$ can be estimated as  
$|{\bm B}^{\textrm{eff}}_{\sigma=\pm}| = (4.3 \mp 2.6) \times 10^{-16}
(\frac{\Omega}{[\rm s^{-1}]}) (\frac{|\bm{\mathcal{E}}|^2}{[\rm V^2/m^2]})$ T.
For $|\bm{\mathcal{E}}|=4$ kV/m and $\Omega=2.2 \times 10^{9}$ s$^{-1}$, 
$|{\bm B}^{\textrm{eff}}_{\sigma=\pm}|$ can reach up to $15\mp 9$ T.
%
%
Then, the induced charge current 
becomes $|\langle \bm{j} \rangle| \simeq 2 \times 10^{6}$A/m$^2$, 
whose current density is much larger than the anomalous Hall current density
due to the chiral anomaly\cite{rf:Sekine15}. 
This giant current density is caused from the giant magnetic field ${\bm
B}^{\textrm{eff}}_\sigma$. 
We would like to point out  that 
$\langle \bm{j} \rangle $ is distinguished from 
the longitudinal \cite{rf:Hosur12} and the
transverse charge current 
\cite{rf:Zyuzin12,rf:Chen13,rf:Burkov15,rf:Sekine15},
since $\langle \bm{j} \rangle $ is parallel to the light traveling
direction and it flows in an opposite direction when 
the chirality of the light is reversed.
%
%

{\bf Floquet state}---
So far, we have assumed that the frequency $\Omega$ of light is much
lower than a scale of the band width.
Now, we consider the opposite case.
In contrast to the lower $\Omega$ case, in which only electrons near the
Fermi surface are influenced by light, the higher frequency light can
affect the whole electrons in valence bands.  

To consider this situation, we adapt
the Floquet method: 
Because $H$ in Eq.(\ref{eq:2-1}) is periodic in $t$, i.e. $H(t)=H(t+2\pi/\Omega)$, 
the wave function of the Sch\"{o}dinger equation
$i\hbar\partial_t\psi(t)=H(t)\psi(t)$ has the form of $\psi(t)=\sum_m \phi_m
e^{-i(\varepsilon+m\hbar\Omega)t/\hbar}$, where the summation is taken for all
integers $m$.
Substituting this form into the Sch\"{o}dinger equation, we have  
the Floquet equation, $\sum_n H_{m,n}\phi_n=(\varepsilon+m\hbar \Omega)\phi_m$, with
$H_{m,n}=(\Omega/2\pi)\int_0^{2\pi/\Omega} dt H(t)e^{i(m-n)\Omega t}+m\hbar \Omega\delta_{m,n}$.
For the Hamiltonian in Eq.(\ref{eq:2-1}), the diagonal term of the Floquet
Hamiltonian is given by $H_{m,m}=H_{\rm Weyl}+V_{\rm imp}+m\hbar \Omega$, and
the off-diagonal ones are 
$H_{m,m+1}=H_{m+1,m}^*=(\Omega/2\pi)\int_0^{2\pi/\Omega}dt H_{\rm
em}e^{-i\Omega t}
= -  \frac{i e  v_\textrm{F}| \bm{\mathcal{E}}| }{2\Omega } \sigma^z(s^x- i\sigma_{\rm{L}} s^y )$ when light is along $z$ axis.
Other off-diagonal terms are identically zero.
Each solution of the Floquet equation gives a periodic steady state.

For large $\Omega$, the diagonal terms are dominant, so one can treat the
off-diagonal ones as a perturbation. 
In the zeroth order, our system is described by 
$H_{0,0}=H_{\rm Weyl}+V_{\rm imp}$, then 
the first non-zero correction in the perturbation theory appears in the
second order as 
$
\frac{1}{\hbar\Omega}\left[H_{0,-1}, H_{0,1}\right].
$
Thus, we obtain the following effective Hamiltonian
\begin{eqnarray}
H_{\rm eff}=H_{\rm Weyl}+V_{\rm imp}+i \sigma^0 \frac{e^2v_{\rm F}^2}{\hbar
 \Omega^3}
(\bm{\mathcal E}\times \bm{\mathcal E}^*)\cdot{\bm s}, 
\label{eq:effectiveHamiltonian}
\end{eqnarray}
by which a periodic steady state of our system is described.

From Eq. (\ref{eq:effectiveHamiltonian}), it is found that the higher
frequency light induces a different effective Zeeman magnetic field, 
\begin{eqnarray}
{\bm B}^{\rm eff}_{\rm Floquet}\equiv
\frac{e^2v_{\rm F}^2}{g\mu_{\rm B}\hbar\Omega^3}
i \sigma^0 (\bm{\mathcal E}\times \bm{\mathcal E}^*).
\label{eq:BeffF}
\end{eqnarray}
Here we note that the physical origin is completely different from  
that in the lower frequency case.
While ${\bm B}^{\rm eff}_\sigma$ in Eq.(\ref{eq:3-4}) originates from a dissipative
process so it depends on $\tau_{{\rm e},\sigma}$,
 ${\bm B}^{\rm eff}_{\rm Floquet}$ in Eq.(\ref{eq:BeffF}) is independent of the impurity scattering. 
Furthermore, the former magnetic field only affects on electrons near
the Fermi surface, but the latter acts on the whole of the band.
Consequently, the resultant phenomena can be different.

We find that no net current $\langle {\bm j}\rangle$ is obtained
by ${\bm B}^{\rm eff}_{\rm Floquet}$:
According to Eq.(\ref{eq:effectiveHamiltonian}), ${\bm B}^{\rm
eff}_{\rm Floquet}$ just provides a uniform Zeeman splitting (or shift) in the
whole band spectrum of the Weyl semimetal, like a static Zeeman field. 
Therefore, in a steady state, electrons fill the band up
to the Fermi energy. In this situation, one can use the same argument in
Ref.\cite{rf:Vazifeh13}, and prove that $\langle {\bm j} \rangle=0$.
Whereas Weyl fermions may have a nonzero spin $\langle
\psi_{\sigma}^{\dagger}{\bm s}\psi_{\sigma}\rangle$ due to the Pauli magnetism of ${\bm B}^{\rm
eff}_{\rm Floquet}$, the current due to the spin-momentum locking is
totally cancelled by the current from the rest of the band.
In other words, no photovoltaic chiral magnetic effect occurs for the
higher frequency light.

It is helpful to regard the frequency $\Omega$ as an energy cutoff 
for the chiral magnetic effect.
For lower $\Omega$, the light can excite only Weyl fermions near the Fermi
surface, and thus the quasi relativistic phenomena like the chiral
magnetic effect may occur. As $\Omega$ increases, electrons in a
lower position of the band can participate in the current, then
eventually, when $\Omega$ is large enough to affect the whole spectrum
of the band, the chiral magnetic effect is completely cancelled.

Instead, for higher $\Omega$, one can
expect the light induced anomalous Hall effect. 
Substituting Eq.(\ref{eq:2-2}) for $H_{\rm Weyl}$ in
Eq.(\ref{eq:effectiveHamiltonian}), one finds that ${\bm B}^{\rm eff}_{\rm
Floquet}$ shifts ${\bm b}$ by $\delta{\bm b}=-(g\mu_{\rm B}/2\hbar
v_{\rm F}) {\bm B}^{\rm eff}_{\rm Floquet}$. 
The change of ${\bm b}$
induces the change of $\theta$-term in the Weyl semimetals \cite{rf:Zyuzin12b}, which
results in  
\begin{eqnarray}
\langle \delta \rho \rangle =\frac{2\alpha
 c\epsilon_0}{\pi}\delta{\bm b}\cdot{\bm B},
\quad
\langle \delta {\bm j} \rangle=-\frac{2\alpha c
 \epsilon_0}{\pi}\delta{\bm b}\times {\bm E},
\end{eqnarray}
in the presence of external magnetic and electric fields, ${\bm B}$ and
${\bm E}$. Here $\alpha$ is the fine structure constant, $c$ is the speed of light, and $\epsilon_0$ is the vacuum permittivity. 
The light induced charge pump $\langle \delta \rho\rangle$ and anomalous
Hall current $\langle \delta {\bm j}\rangle$ have been discussed
recently in Refs.\cite{rf:Oka15,rf:Chan15,rf:oka09}. 






{\bf Conclusion}---
We theoretically predict photovoltaic chiral magnetic effect, which is induced by the effective magnetic field due to circularly polarized light.
In the low light frequency regime, 
the effective magnetic field affects only fermions near the Fermi surface. 
As a result, the effective magnetic field plays the role to trigger a
finite spin polarization of Weyl fermions and drive the finite charge current in
Eq.(\ref{eq:current}). 
On the other hand, in the high frequency regime, the Floquet quasi
steady state is realized.
The circularly polarized light induces the effective magnetic field in
Eq.(\ref{eq:BeffF}),  
which is completely different from that in the lower frequency regime. 
The magnetic field in the high frequency regime behaves like the Zeeman
field and shifts the whole 
band structure. 
The current of Weyl fermions are completely cancelled by other band
contribution.
Our photovoltaic chiral magnetic effect, which drastically
depends on the light frequency,
realizes the chiral magnetic effect in condensed matter
physics.

\section*{Acknowledgments}
The authors acknowledge the fruitful discussion with K. T. Law and W. Y. He.
This work was supported by Grants-in-Aid for the Core Research for Evolutional Science and Technology (CREST) of the Japan Science, for Japan Society for the Promotion of Science (JSPS) Fellows, 
for Topological Materials Science (TMS) (Grant No.15H05855, No.15H05853), 
for Scientific Research B (Grant No.25287085, No.15H03686), 
and 
for Challenging Exploratory Research (Grant No.15K13498).

\end{document}